%% file: paper.tex
\newif\ifdraft
\draftfalse

\newif\iftwocolumn
\twocolumnfalse

\ifdraft
  \documentclass[12pt, twocolumn]{article}
\else
  \documentclass[twocolumn]{article}
\fi

\usepackage[margin=1in]{geometry}
\usepackage[parfill]{parskip}
\usepackage[utf8]{inputenc}
\usepackage{hyperref}

\usepackage[square,numbers]{natbib}

\usepackage{amsmath,amssymb,amsfonts,amsthm}

\usepackage{graphicx}

\usepackage[export]{adjustbox}
\graphicspath{ {images/} }
\let\oldincludegraphics\includegraphics
\renewcommand{\includegraphics}[2][]{%
  \oldincludegraphics[#1,max width=\linewidth]{#2}}

\usepackage{abstract}

\ifdraft
  \usepackage{lineno}
  \usepackage{xcolor}
  
  \linenumbers
\fi

\title{\textbf{Time-warped Trials}}

\author{Eric Easthope \\
        \small{University of British Columbia} \\
        \small{Department of Electrical and Computer Engineering} \\
        \small{Vancouver, British Columbia, Canada}
        }

\date{}

\begin{document}

\onecolumn
\maketitle

\begin{abstract}
  \normalsize
  \input{./build/abstract}
\end{abstract}

\iftwocolumn
  \twocolumn
\else
  \onecolumn
\fi

\input{./build/body}

\bibliographystyle{plain}
\bibliography{paper}

\end{document}

%% file: build/abstract.tex
I outline a signal resampling strategy for aligning event times between
time series trials in contexts where significant event times like onsets
and offsets vary between trials. These variations prevent direct
comparisons of trials in practical contexts as comparisons require
equal-length time series \citep{Salari2019}. Algorithms like dynamic
time warping help us quantify these variations locally but do not apply
well to continuous transformations of time series signals without
interpolating or downsampling to add or remove samples
\citep{Sampling2014, DTWBad2004}. I show that with consideration for
padding and sampling frequency that sinc interpolation is sufficient to
resample parts of trial intervals to produce equal-length time-locked
trials that correlate to and strongly approximate their unwarped
counterparts with minimal interpolation effects. Specifically I show
that interpolation effects can be minimized by oversampling, selectively
interpolating mis-aligned parts of trials with respect to mean
mis-aligned event lengths, and interpolating mis-aligned events with
sufficient zero-padding. Interpolated signals then have a bandlimit well
below the Nyquist frequency and satisfy the Nyquist-Shannon sampling
theorem ensuring perfect reconstructions, and I find that I can track
and potentially counteract resampling effects on signal energy
quantities.

%% file: build/body.tex
\hypertarget{background}{%
\section{Background}\label{background}}

Signal events are often windowed in the time domain as trials marked by
the start and end of some repeated signal of interest. Ideally these
trials are time-aligned such that trial lengths are fixed. However due
to context, experimental circumstance, and/or human error these trial
start and end times can vary slightly. This is more pronounced in trials
containing other significant event times such as transition times
between parts. As trials are commonly represented digitally by sequences
of numbers this makes it challenging if not impossible to directly
compare trials of different lengths through simple operations like
averaging. Operations between variable-length sequences are generally
poorly defined or arbitrary and workarounds are required to time-lock
events.

Techniques like \emph{dynamic time warping} (DTW) \citep{DTW1959} and
\emph{nonlinear mixed effects modelling}
\citep{NonlinearMixedEffects2014} help us quantify and adjust for
differences between variable-length sequences but do not apply well to
continuous time series without interpolating or downsampling sequences
to add or remove samples \citep{Sampling2014, DTWBad2004}. DTW is also
designed with pairs of discrete sequences in mind and is generally
applied only to pairs of sequences \citep{DTWPairs1996}. Time complexity
and DTW susceptibility to overfitting to noise \citep{DTWOverfit2016}
are limiting for trial data processing and heuristics are required to
operate on larger sets of sequences simultaneously
\citep{DTWGlobal2011}.

I explore a simpler way to make minimal, quantifiable adjustments to
variable-length time series sequences to enable direct comparisons of
trial time series quantities.

\hypertarget{method}{%
\section{Method}\label{method}}

Sinc interpolation is a known resampling strategy for downsampling or
upsampling signals from one sampling frequency to another and is a good
candidate for time-locking event sequences.

Resampling a single time interval per trial is somewhat trivial and
already well-defined in existing literature and presumes that the
interval we wish to time-align starts and stops at the beginning and end
of the trial. We are interested in non-trivial cases where one or more
significant event times like onsets and offsets vary between trials and
need to be time-locked with respect to each other.

\begin{figure}
\centering
\includegraphics{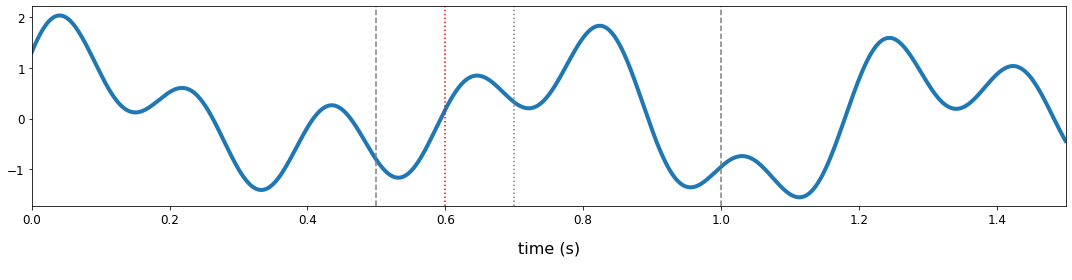}
\caption{A non-periodic signal sampled at \(f_{samp} = 2048\,\text{Hz}\)
composed of frequencies \(f_1 = 5/\pi\) and \(f_2 = 5/2\). Significant
event times are annotated (black and red).\label{figure-1}}
\end{figure}

For purposes of demonstration I construct a sample trial as a
non-periodic time series sampled at \(f_{samp} = 2048\,\text{Hz}\) with
three significant event times shown in Figure \ref{figure-1} composed of
two frequencies \(f_1 = 5/\pi\) and \(f_2 = 5/2\). Evidently
\(f_1, f_2\) are mutually non-divisible by one another and are well
below the Nyquist frequency as
\(f_{nyq} = f_{samp}/2 = 1024\,\text{Hz} \gg f_1, f_2\) so
Nyquist-Shannon sampling theorem says that we will get perfect signal
reconstructions through sinc interpolation.

The simplest case of interest is a trial made of two time intervals
bisected by a single event time. This arises in situations where a
single trial event has some variance in its onset (or offset) time such
as during a button press or the start of some vocalization in response
to trial stimuli. We also see this in situations where non-event time
intervals precede and succeed two intervals of interest such as in
Figure \ref{figure-1}.

In both situations the same resampling strategy applies and time-locking
events across trials while preserving trial length requires that one
time region \(T_1\) is contracted with respect to the expansion of the
other \(T_2\). We partition each trial using indices determined by event
onset and offset times and resample the middle two time intervals.

Signals from \(T_1, T_2\) are padded with \(\ell_{pad}\) values from
preceding and succeeding signals adjacent to \(T_1, T_2\) to prevent
``ringing'' artifacts observed when signal lengths are too short for
sinc interpolation. How I choose \(\ell_{pad}\) is somewhat arbitrary
before I justify it quantitatively so I qualitatively choose a large
enough \(\ell_{pad}\) value so that ringing artifacts visually occur
away from \(T_1, T_2\). In my discussion I quantify how small
\(\ell_{pad}\) can be without detriment to resampling accuracy.
Qualitatively it seems that padding with
\(\ell_{pad} \approx 10\% \times f_{samp}\) points per side per trial is
more than sufficient to evade ringing artifacts in this construction.

I choose how long resampled regions \(T_1', T_2'\) should be based on
context and compute the rescaling ratios \(r_1, r_2\) for each interval.
Contracted intervals correspond to \(T_i > T_i'\) and \(r_i > 1 > 0\)
whereas expanded intervals correspond to \(T_i < T_i'\) and
\(0 < r_i < 1\). To resample and accordingly warp each trial with
respect to time I apply sinc interpolation with a windowed sinc filter
to each trial signal from \(T_1, T_2\). Then I truncate resampled
signals in \(T_1, T_2\) by \(\ell_{pad, 1}, \ell_{pad, 2}\) respectively
and concatenate these truncated signals with non-scaled trial intervals
to produce time-locked trials. Different outcomes based on contracting
or expanding \(T_1, T_2\) are highlighted in Figure 2.

\begin{figure}
\centering
\includegraphics{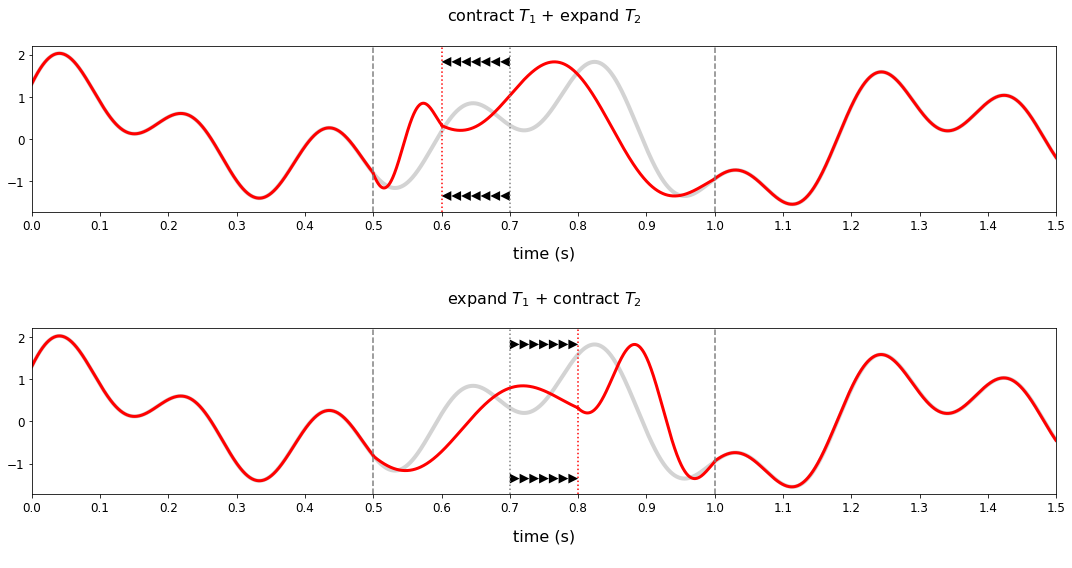}
\caption{Different outcomes based on resampling to contract or expand
\(T_1, T_2\). Unwarped signals are shown in grey.\label{figure-2}}
\end{figure}

\hypertarget{discussion}{%
\section{Discussion}\label{discussion}}

\hypertarget{time-locked-trials-strongly-approximate-unwarped-trials}{%
\subsection{Time-locked trials strongly approximate unwarped
trials}\label{time-locked-trials-strongly-approximate-unwarped-trials}}

\begin{figure}
\centering
\includegraphics{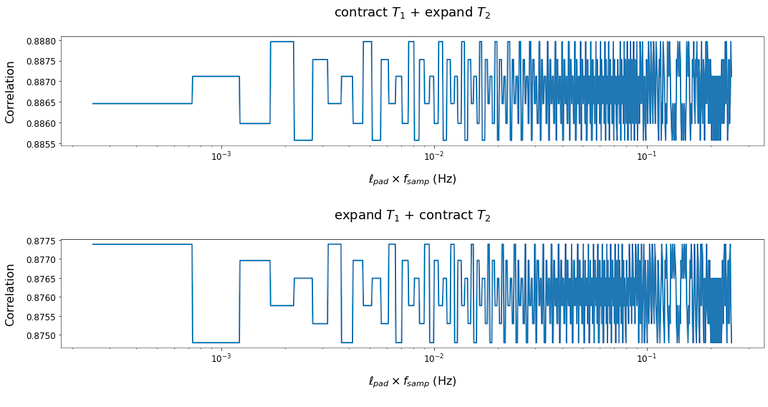}
\caption{Warped trial correlations (y-axis) with respect to padding
choice. Padding is expressed in terms of sampling frequency \(f_{samp}\)
(x-axis) for expanded and contracted intervals \(T_1, T_2\)
(rows).\label{figure-3a}}
\end{figure}

Visibly the warped (sinc-interpolated) signals have more than reasonable
correlation (\(r \approx 0.85\), Figure \ref{figure-3a}) to their
unwarped counterparts with the possible exception of interval endpoints
where resampling by contraction or expansion creates non-smooth (but
piecewise continuous) curves at event onset, transition, and offset
times. Appropriately chosen peak finding and linear interpolation might
fix some of these artifacts.

\begin{figure}
\centering
\includegraphics{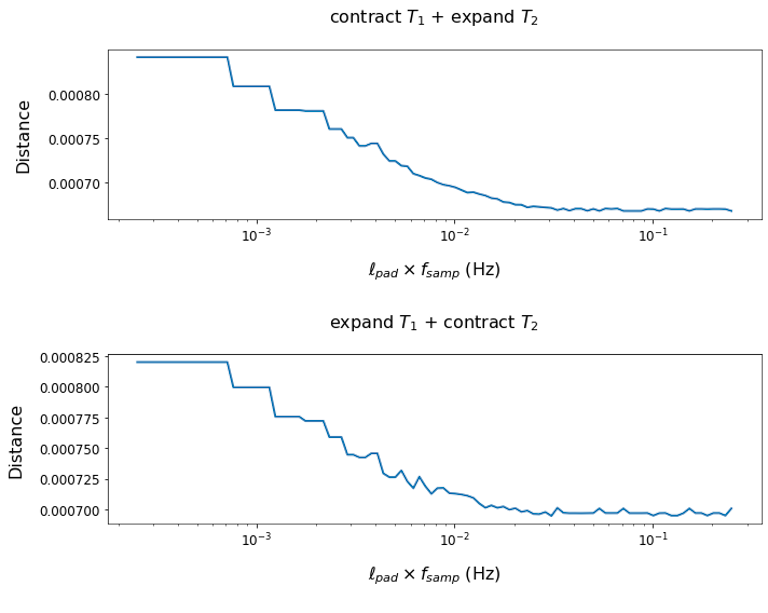}
\caption{Normalized Euclidean DTW distances between warped and unwarped
trials with respect to padding choice. Padding is expressed in terms of
sampling frequency \(f_{samp}\) (x-axis) for expanded and contracted
intervals \(T_1, T_2\) (rows).\label{figure-3b}}
\end{figure}

\begin{figure}
\centering
\includegraphics{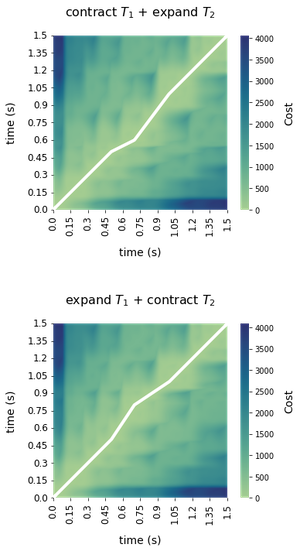}
\caption{Effective cost matrix of using DTW to transform between warped
signals and their unwarped counterparts in intervals \(T_1, T_2\). The
minimal ``warping path'' is highlighted showing near-zero cost to
transform between warped and unwarped signals.\label{figure-4}}
\end{figure}

I go further and use Euclidean DTW distance as a proxy metric for
interpolation quality and compute the effective ``distance'' between
warped and unwarped signals. I compare these with respect to sampling
frequency in Figure \ref{figure-3b}. Using normalized Euclidean
distances, warped signals are more than 99\% similar to their unwarped
counterparts and the ``cost'' of warping and unwarping trials in Figure
\ref{figure-4} shows that trial transformations are nearly trivial to do
and conversely to undo predictably too. Retaining strong if not perfect
correspondence in non-scaled intervals and overall trial duration
highlights the importance of rescaling parts of trials in a
complementary way so as to preserve total trial lengths and correlations
in fixed-length intervals outside of \(T_1, T_2\).

\hypertarget{padding-choice-depends-on-f_samp}{%
\subsection{\texorpdfstring{Padding choice depends on
\(f_{samp}\)}{Padding choice depends on f\_\{samp\}}}\label{padding-choice-depends-on-f_samp}}

\begin{figure}
\centering
\includegraphics{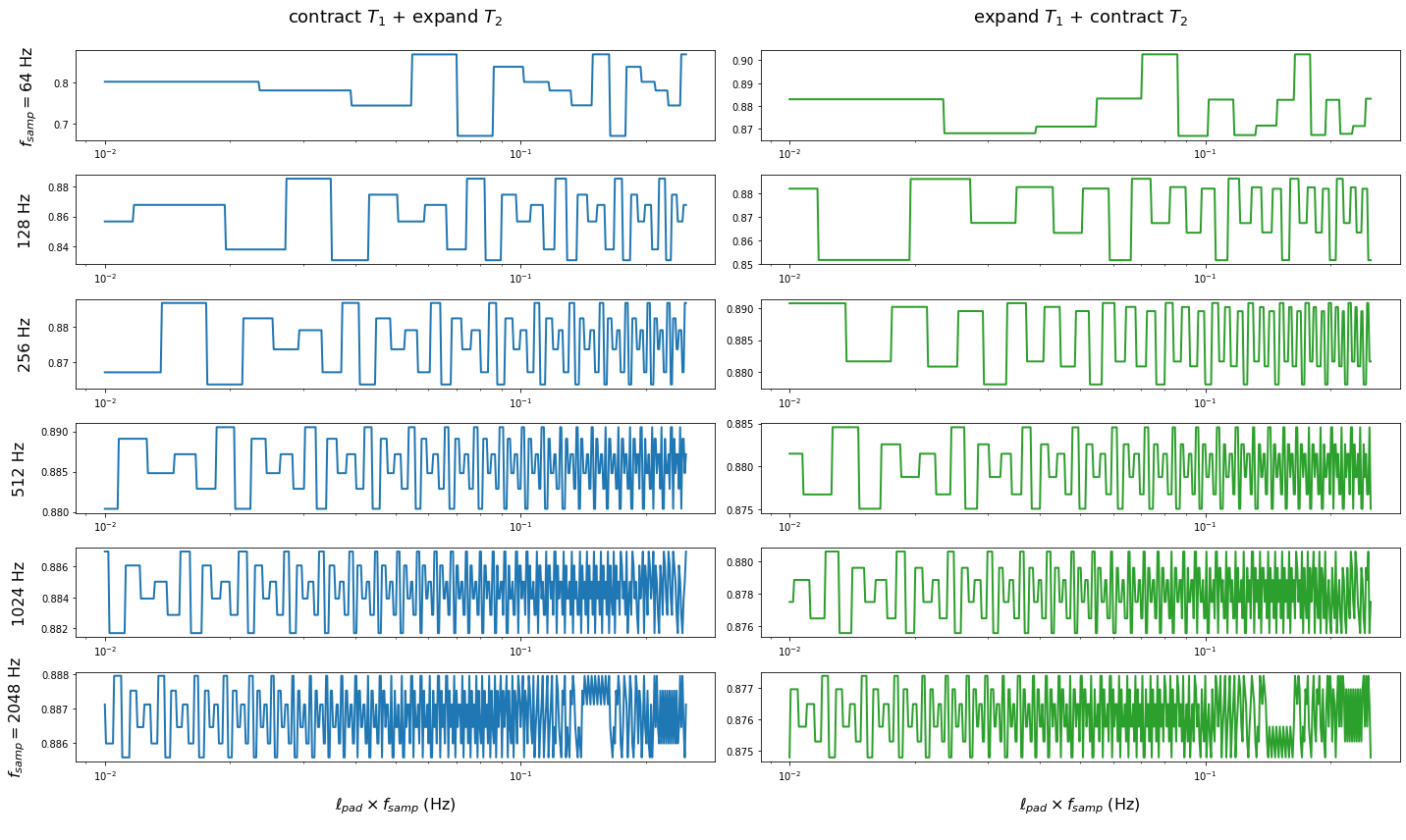}
\caption{Warped trial correlations (y-axis) with respect to padding
choice. Padding is expressed in terms of sampling frequency \(f_{samp}\)
(x-axis, rows) for expanded and contracted intervals \(T_1, T_2\)
(columns).\label{figure-5a}}
\end{figure}

\begin{figure}
\centering
\includegraphics{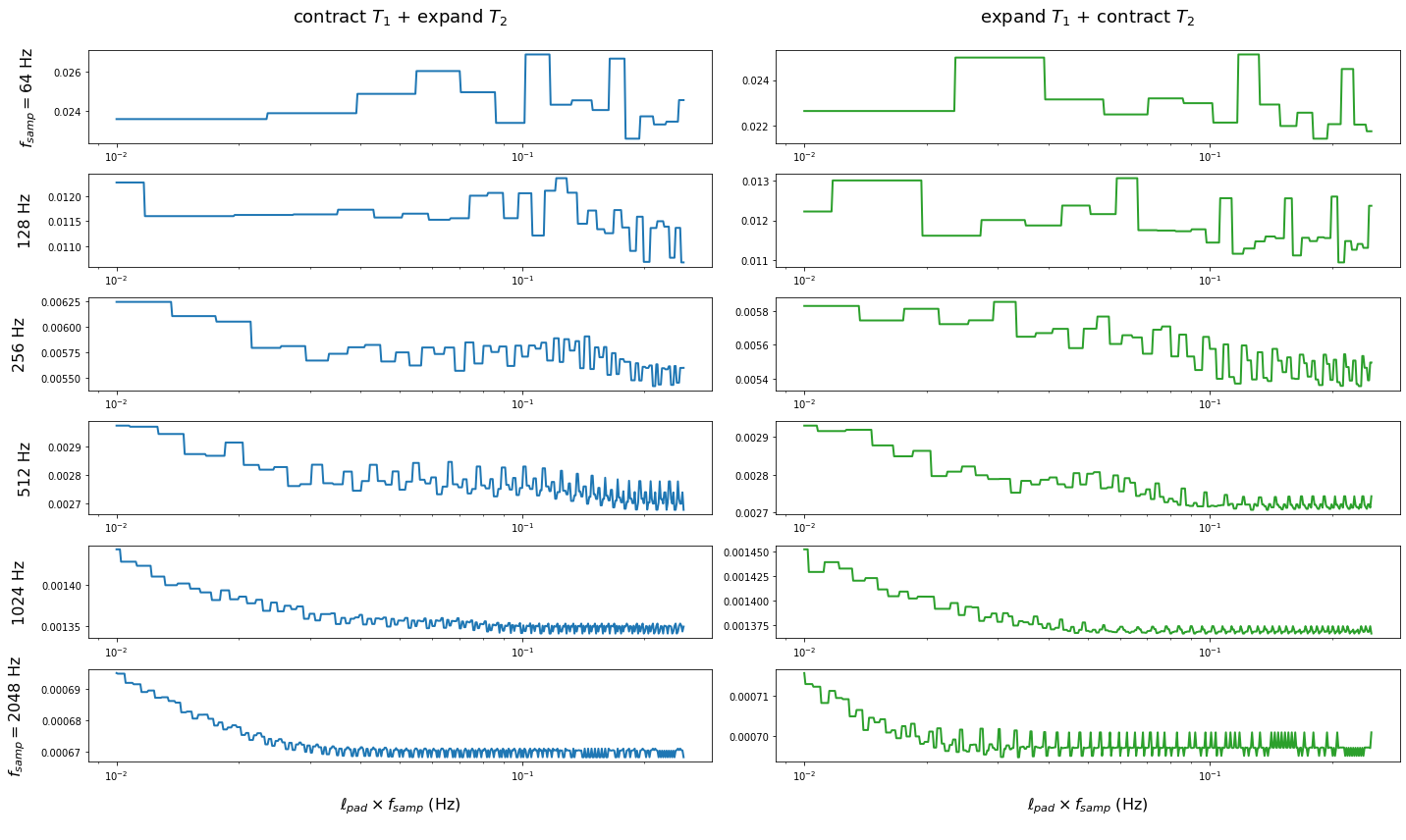}
\caption{Normalized Euclidean DTW distances between warped and unwarped
trials with respect to padding choice. Padding is expressed in terms of
sampling frequency \(f_{samp}\) (x-axis, rows) for expanded and
contracted intervals \(T_1, T_2\) (columns).\label{figure-5b}}
\end{figure}

I estimate the padding sufficient to warp each interval without
significant interpolation artifacts in Figures \ref{figure-5a} and
\ref{figure-5b} and we see that \(\ell_{pad}\) can go as low as
\(\ell_{pad} \approx 10^{-2}\) before interpolation artifacts like
ringing effects at the endpoints of ``short'' signals degrade
correlation and DTW distances. Sufficiently padded intervals
\(\left(\ell_{pad} \ge 10^{-1}\right)\) remain reasonably correlated and
Nyquist-Shannon sampling theorem says that these correlations will
remain at higher sampling frequencies. Decreasing \(f_{samp}\) we see
little to no effect on correlations to as low as
\(1/32 \times f_{samp}\) in Figure \ref{figure-5a}. Interpolated signals
lose 99\% similarity as low as \(1/8 \times f_{samp}\) in Figure
\ref{figure-5b} suggesting a significant factor of sampling overhead
before sinc interpolation fails to reconstruct similar time-warped
signals.

\hypertarget{signal-energy-scales-linearly-with-interpolation}{%
\subsection{Signal energy scales linearly with
interpolation}\label{signal-energy-scales-linearly-with-interpolation}}

From a change of variables we see that the energy of any non-periodic
signal \(f\) when warped to \(f'\) over some uniformly-spaced finite
time domain \(t \in [0, L]\) is scaled by at most a linear factor
\(r_i\) by resampling to another uniformly-spaced finite time domain
\(r_i t = T \in [0, r_iL]\):

\[r_i \cdot E_{f} = r_i \cdot \langle{ f(t), f(t)\rangle} \equiv \langle{ f(r_i t), f(r_i t)\rangle} = \langle{ f(T), f(T)\rangle} = E_{f'}.\]

Correspondingly signal powers \(P_i = \int |f(t)|^2 dt\) per \(T_i\)
should scale linearly with sinc interpolation up to a factor of
numerical floating point error. This makes changes in energy easy to
track and account for numerically and in fact \(r_i\) is the linear
energy scaling factor per \(T_i\).

\hypertarget{conclusion}{%
\section{Conclusion}\label{conclusion}}

I show that sinc interpolation is sufficient as a resampling strategy to
warp and produce equal-length time-locked trials that correlate
reasonably well to their unwarped counterparts without unwanted
interpolation effects. Specifically I show that interpolation effects
can be minimized by oversampling, selectively interpolating mis-aligned
parts of trials with respect to mean mis-aligned event lengths, and
interpolating mis-aligned events with sufficient zero-padding. I
estimate the minimum interpolation padding \(\ell_{pad}\) to produce
99\% similar reconstructions with respect to sampling frequency
\(f_{samp}\) and approximate how low \(f_{samp}\) and correspondingly
how low \(\ell_{pad}\) can be before interpolation degrades under
sampling and bandwidth constraints. Finally I find that I can track
resampling effects on signal energy quantities.